# Magnetic dead layers in $La_{0.7}Sr_{0.3}MnO_3$ revisited

S. B. Porter[1], M. Venkatesan[1], P. Dunne[2], B. Doudin[2], K. Rode[1], and J. M. D. Coey[1]

[1]CRANN and School of Physics, Trinity College, Dublin 2, Ireland

[2]IPCMS, 67034 Strasbourg, France

*Abstract*—The magnetic dead layers in films a few nanometers thick are investigated for $La_{0.7}Sr_{0.3}MnO_3$ on (001)-oriented $SrTiO_3$ (STO), $LaAlO_3$ (LAO) and $(LaAlO_3)_{0.3}(Sr_2TaAlO_6)_{0.7}$ (LSAT) substrates. An anomalous moment found to persist above the Curie temperature of the $La_{0.7}Sr_{0.3}MnO_3$ films is not attributed to the films, but to oxygen vacancies at or near the surface of the substrate. The contribution to the moment from the substrate is as high as 20 $\mu_B/nm^2$ in the case of STO or LSAT. The effect is increased by adding an STO cap layer. Taking this *d*-zero magnetism into account, extrapolated magnetic dead layer thicknesses of 0.8 nm, 1.5 nm and 3.0 nm are found for the manganite films grown on LSAT, STO and LAO substrates, respectively. An STO cap layer eliminates the LSMO dead layer.

## I. INTRODUCTION

MIXED valence manganites with optimal cation doping have been the subject of numerous studies ever since their discovery in 1950, when they were the first known ferromagnetic oxides [1]. Exhibiting Zener double exchange, half-metallicity, a wide tunable range of Curie temperatures $T_C$ and colossal magnetoresistance near $T_C$ in thin films, the mixed valence manganite families have interesting magnetic and electronic properties and potential for applications [2,3], not least as electrodes for thin-film oxide and organic spin electronics [4-7]. The general formula of the perovskite manganites is $(A_{1-x}B_x)MnO_3$, where A and B are rare-earth and alkaline-earth or $Pb^{2+}$ cations respectively. The manganese cations, formally $Mn^{3+}$ and $Mn^{4+}$, are octahedrally coordinated by oxygen, and the octahedra tend to exhibit Jahn-Teller distortion when occupied by $Mn^{3+}$ $3d^4(t_{2g}^3e_g^1)$. The compositional phase diagrams exhibit phases with a wide variety of electronic structure, magnetic, charge and orbital order. Magneto-transport properties of the conducting mixed-valence manganites are strongly influenced by double exchange, where $e_g$ electron hopping between adjacent manganese sites is only permitted when the spins of the $t_{2g}^3$ $Mn^{4+}$ cores on the two sites are aligned parallel. 'Optimally-doped' manganites with $x \approx 0.3$ are the best metals, with the highest Curie temperatures. The magnetic properties of thin films have recently been shown to be influenced by tilting of the network of interlinked, corner-sharing octahedra that make up the perovskite structure [8-10].

Two disadvantages of these oxides, limiting their usefulness, are the relatively low Curie temperatures, which do not exceed the 380 K achieved for $La_{0.7}Sr_{0.3}MnO_3$ (LSMO), and the tendency for thin films and nanoparticles to form a magnetic 'dead layer', which is a region at the surface or interface with the substrate where the ferromagnetic properties of the manganite are greatly diminished or modified. An early investigation of the properties of LSMO by spin-resolved photoemission spectroscopy by Park *et al.* [11] showed that the temperature dependence of the magnetization of the surface differed significantly from that of the bulk.

An electric dead layer in LSMO has been characterized in films grown on $LaAlO_3$, $NdGaO_3$ and MgO substrates [12,13]. The dead layer in these cases is attributed either to structural disorder or to chemical alteration of the interfacial layer during deposition, and it turns out to be thicker than the magnetic dead layer.

More recent studies by Boschker *et al.* [14] have suggested that very thin, 2 nm LSMO films form a high-temperature insulating phase with a Curie temperature of 560 K. The same group has reported moments of 4.0 $\mu_B$/Mn, significantly in excess of the spin-only moment of 3.7 $\mu_B$ expected from the formula $(4 - x)\mu_B$ [15]. Very recently, Liao *et al.* [10] have shown that magnetic dead layers in LSMO films grown on (110) MdGaO$_3$ can be eliminated by an STO capping layer.

In this work, we examine the thickness of the magnetic dead layer in LSMO on STO, LAO and LSAT (001) substrates and investigate the influence of the substrates on the overall magnetic moment. These substrates have lattice parameters of 391, 379 and 387 pm, respectively, compared with 390 pm for LSMO. Any magnetism of the substrate [16] could be especially important when measuring ultrathin films. Furthermore, we also examine the effect of an STO capping layer on films deposited on STO. The electric dead layer thickness is investigated for the same substrates. It has been suggested in [17] that the magnetic dead layer exhibits antiferromagnetism, and an insulating Jahn-Teller distorted phase layer is thought to lie at the interface or surface. While we agree that the LSMO magnetism is reduced by weakened double exchange in thin films, our conclusion is that *d*-zero magnetism [18] associated with STO thin films and substrate interfaces [19] is the source of much of the anomalous behavior attributed to LSMO films.

## II. Experiments

The thin LSMO films used in these experiments were grown on polished single-crystal (001) substrates of LSAT, LAO or STO by pulsed laser deposition. The target was prepared by precipitation, drying and calcination of a stoichiometric aqueous solution of the nitrates of La, Sr, and Mn, followed by pressing at 150 bar and sintering at 1050 °C. A KrF excimer laser ($\lambda = 248$ nm) was used to ablate the target at a repetition rate of 3 Hz and a fluence of 1.3 Jcm$^{-2}$, while the sample was maintained at a temperature of 700 °C for the duration of the deposition. A high oxygen pressure of 0.3 mbar was maintained in the deposition chamber during the deposition and during cooling to minimize the presence of oxygen vacancies in the films.

The growth of the films was monitored using in situ reflection high-energy electron diffraction (RHEED), which allowed for precise determination of thickness from the oscillations in intensity corresponding to the growth of individual monolayers. RHEED also gave an indication of the quality of film surface by the shape of the diffraction pattern, despite significant diffusion of the electron beam by the high oxygen pressure present in the chamber.

Samples were structurally characterized by X-ray diffraction and X-ray reflectometry using a standard (line source) laboratory diffractometer (Philips Panalytical X'Pert Pro) to determine the out-of-plane lattice parameter and film thickness.

Magnetic characterization was carried out using a SQUID magnetometer (Quantum Design MPMS) at 300 K and 4 K. Electrical characterization was performed with a Keithley 2400 source meter in the 4-wire van der Pauw geometry using National Instruments Labview software.

## III. Results

With all three substrates adopted in this study, a bulk-like magnetization of 3.6 ± 0.1 $\mu_B$/Mn and Curie temperature of 360 K can be achieved at sufficient thickness. The magnetization curves of films of similar thickness exhibiting these properties are shown in Fig 1. Table I lists the sample thicknesses and the magnetization value based on this thickness in columns 2 and 4, respectively. The Curie temperatures determined from thermal scans in a field of 100 mT, as shown for three films in Fig. 2, are listed in column

5. All films, except the thinnest ones on LAO and LSAT, are ferromagnetic.

The scans in Fig. 2 all exhibit a Curie temperature, but a significant moment persists at high temperatures, well above the Curie temperature of the LSMO. This temperature-independent contribution is most apparent in the thinnest films and it appears to be coming from the substrate. A comparable 'ferromagnetic-like' moment is seen for blank STO, LAO and LSAT substrates, especially when they have been exposed to normal deposition conditions for 30 minutes, as can be seen in Fig 3. These observations agree with results of an earlier study of residual magnetism in common substrates, and the effect of heating them in a reduced oxygen pressure [16]. A detailed study of STO crystals from different sources [19] established that this $d$-zero magnetism is related to the presence of oxygen vacancies at or near the STO surface. It is practically anhysteretic and independent of temperature, at least below 380 K. The LSMO magnetization, corrected for this substrate contribution, shown in column 4 of the Table, increases with film thickness.

This corrected magnetization multiplied by the thickness of the films is then plotted against film thickness in Fig 4. Linear extrapolation to the $x$-axis then gives the dead layer thickness directly as the intercept. The magnetic dead-layer thicknesses for LSAT, STO and LAO substrates are found to be 0.8 nm, 1.5 nm and 3.0 nm, respectively**.**

The sheet conductance of sets of films grown on LSAT, STO and LAO are plotted in Fig. 5, where the inferred electric dead layer thicknesses are found to be about 3 nm, 4 nm and 6 nm respectively, although it is more difficult to determine the electrical dead-layer thickness accurately.

Finally we show the effect of an STO cap layer on two thin films of LSMO on STO in Fig. 6. The Curie temperature does not increase appreciably, but the magnitude of the signal above $T_C$ is enhanced compared with films of similar thickness in Fig. 2. It appears that the surface and interface of the STO capping layer are also contributing a 'quasi-ferromagnetic' signal, like that of the STO substrate itself.

## IV. DISCUSSION

It is now clear that a significant contribution to the moment measured in ultrathin films of LSMO comes from the substrate. This may provide an explanation for anomalously high magnetic moments of 4.0 $\mu_B$/Mn reported for some ultrathin films of LSMO in [14]. This substrate moment leads to potential underestimation of the dead layer thickness when it is not taken into account.

Films grown on STO and LAO with thickness less than 4 nm are found to be insulating, which is in agreement with recent literature [20]. The electric dead layer could be attributed to the existence of a static Jahn-Teller distorted interfacial layer, which favors the occupation of one of the the Mn $e_g$ ($3z^2$-$r^2$) or ($x^2$-$y^2$) $d$-orbitals, depending on whether strain imposed by the substrate is compressive or tensile [17]. Another consequence of the 'quasi-ferromagnetic' substrate signal is that the apparently high Curie temperature can be mistakenly attributed to a thin, insulating manganite film. This may explain high Curie temperature presented in [14].

Comparing Figures 2 and 6, the STO-capped LSMO films show significantly enhanced magnetic moments. This corresponds to a reduction of the dead layer thickness to zero, within the experimental error of 0.3 nm.

A potential consequence of an interface between an insulating LSMO film and nonpolar STO is charge transfer to the STO [21]. The (001) LSMO has a layer charge of ± 0.7e per unit cell, whereas the (100) STO is non-polar. However, the band gap of LSMO is less than that of STO, so the charge transfer is towards the LSMO side of the interface, as for LaMnO₃/STO [22], rather than the STO side as for LAO/STO. Mobile oxygen vacancies may also be involved in the interface reconstruction [23]. Electron redistribution in the LSMO film does not, however, change the average manganese moment as long as the film remains ferromagnetic, but weakened double exchange leading to increasingly dominant antiferromagnetic Mn-Mn interactions will produce spin canting and reduce the average Mn moment, as seen in column 4 of Table 1.

## V. Conclusion

We have confirmed that there is a magnetic dead layer $2 - 8$ unit cells thick in very thin LSMO films, that depends on the substrates on which they are usually grown. Defect-related 'quasi-ferromagnetic' signals arise from the substrates when they are processed in the conditions required to create good-quality LSMO thin films, and this *d-zero* magnetism exhibits little or no temperature-dependence. It is likely responsible for the anomalously high Curie temperatures [14] or Mn moments [15] that have been attributed to some very thin films of LSMO. Further studies of the manganite films by X-ray absorption and X-ray dichroism should permit the separate characterization of the dead layers at the upper surface and substrate interface of these films, and confirm our finding that the dead layer is largely associated with the free surface of the films.

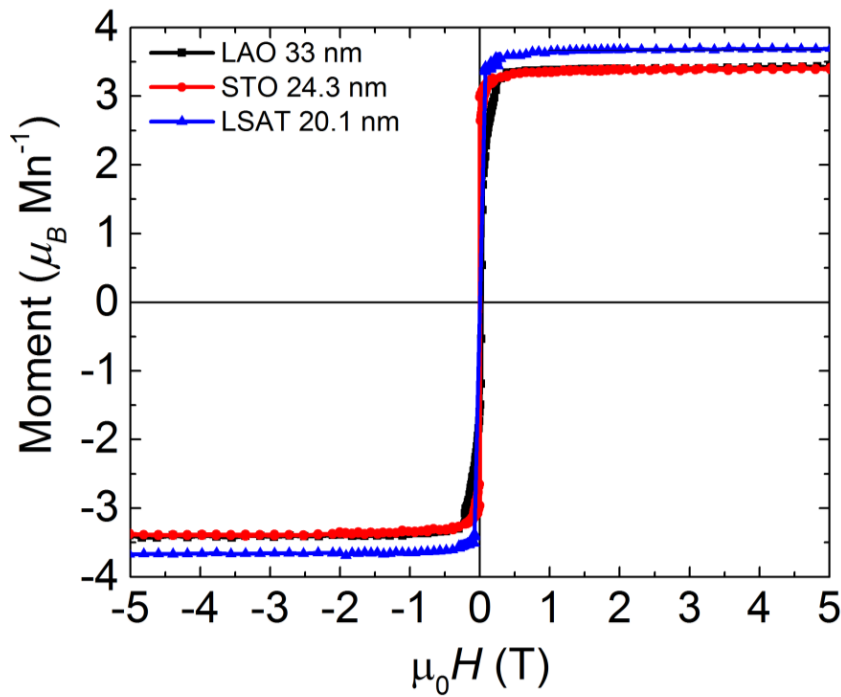

Fig 1. Low-temperature magnetization curves of LSMO magnetization curves on STO, LAO and LSAT substrates. The magnetization of all three films approach the bulk ferromagnetic spin-only-like value of 3.7 μ$_B$/Mn at thicknesses above 20 nm.

| Substrate | Thickness (nm) | $M_{Low-T}$[a] (kAm$^{-1}$) | $M_{corr}$ (kAm$^{-1}$) | $T_C$ (K) |
|---|---|---|---|---|
| LSAT | 1.1 | 631 | 307 | 0 |
| LSAT | 2.2 | 625 | 466 | 200 |
| LSAT | 4.5 | 501 | 421 | 310 |
| LSAT | 8.1 | 532 | 488 | 330 |
| LSAT | 20.1 | 585 | 567 | 360 |
| STO | 1.9 | 440 | 162 | 100 |
| STO | 2.9 | 400 | 217 | 225 |
| STO | 4.3 | 444 | 318 | 290 |
| STO | 5.9 | 558 | 466 | 310 |
| STO | 11.9 | 607 | 561 | 340 |
| STO | 24.3 | 531 | 509 | 340 |
| STO | 48.0 | 571 | 560 | 345 |
| LAO | 1.9 | 97 | <10 | 0 |
| LAO | 6.1 | 305 | 264 | 260 |
| LAO | 8.6 | 356 | 327 | 270 |
| LAO | 14.9 | 472 | 456 | 340 |
| LAO | 32.9 | 541 | 533 | 360 |
| LAO | 48.0 | 553 | 548 | 360 |

[a]Due to the presence of a significant amount of paramagnetic impurities, in the LSAT, films grown on LSAT substrates were measured at 50 K. The others were measured at 4 K.

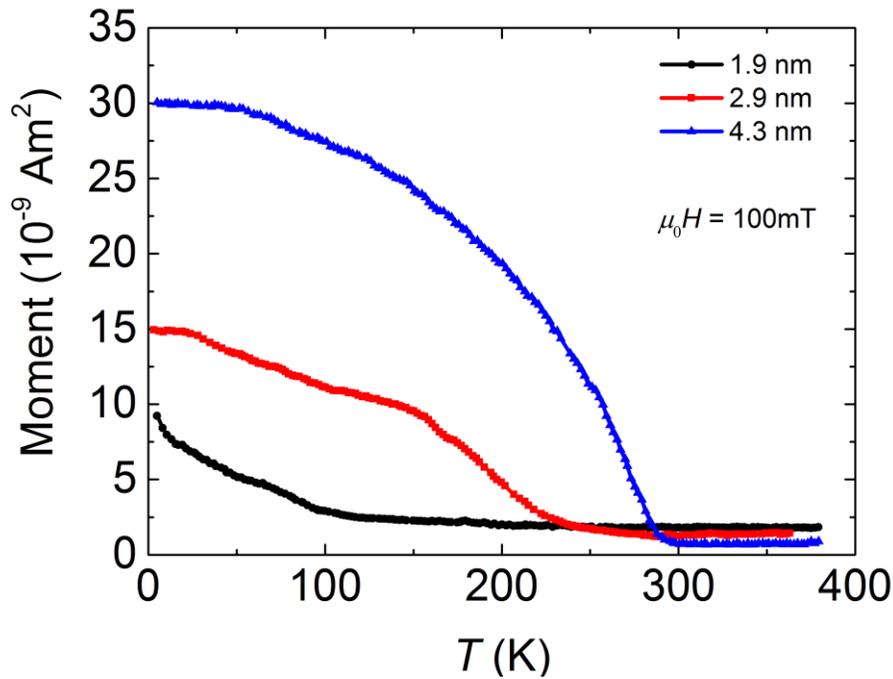

Fig 2. Thermal scans in 100 mT of the magnetization of three thin films of LSMO on STO with different thicknesses of STO capping layers.

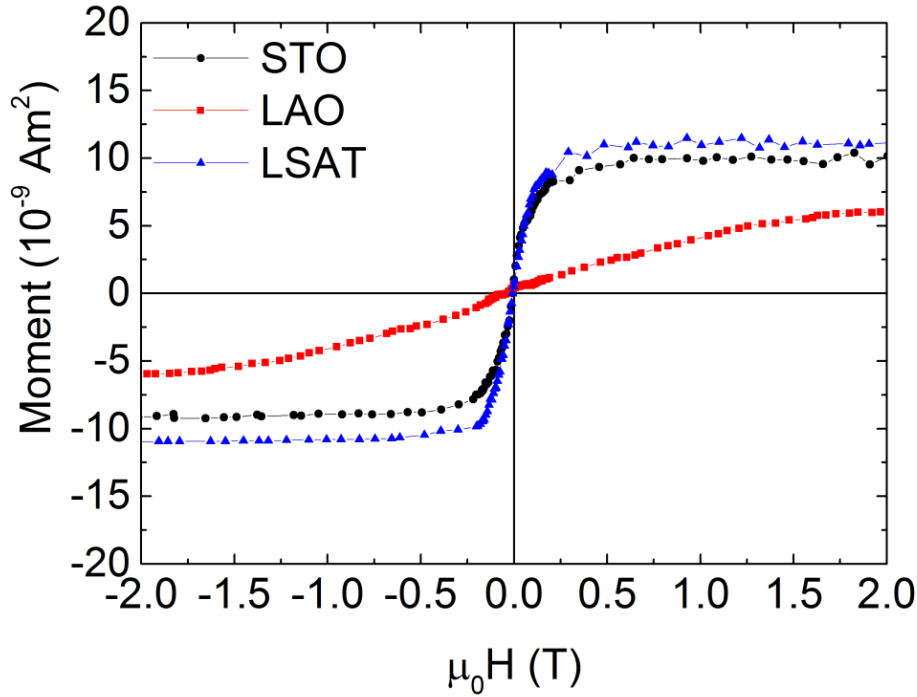

Fig 3. Room-temperature magnetization curves of STO, LAO and LSAT substrates treated in the PLD deposition chamber at 700°C in an oxygen pressure of 0.3 mbar.

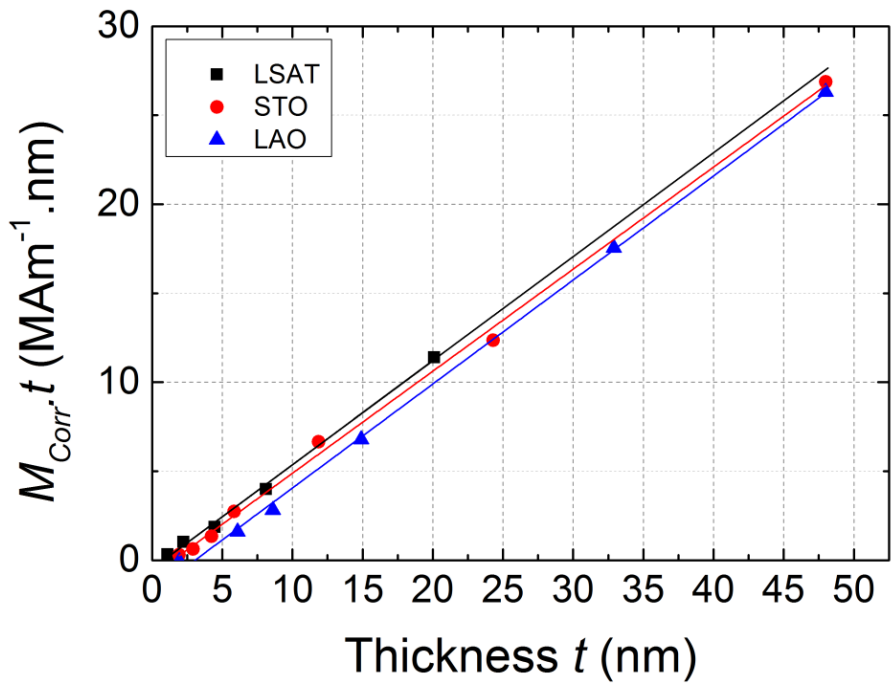

Fig 4. Plot to determine the magnetic dead-layer thicknesses of LSMO films on different substrates. The intercept of the linear fit on the *t*-axis is taken to be the dead layer thickness.

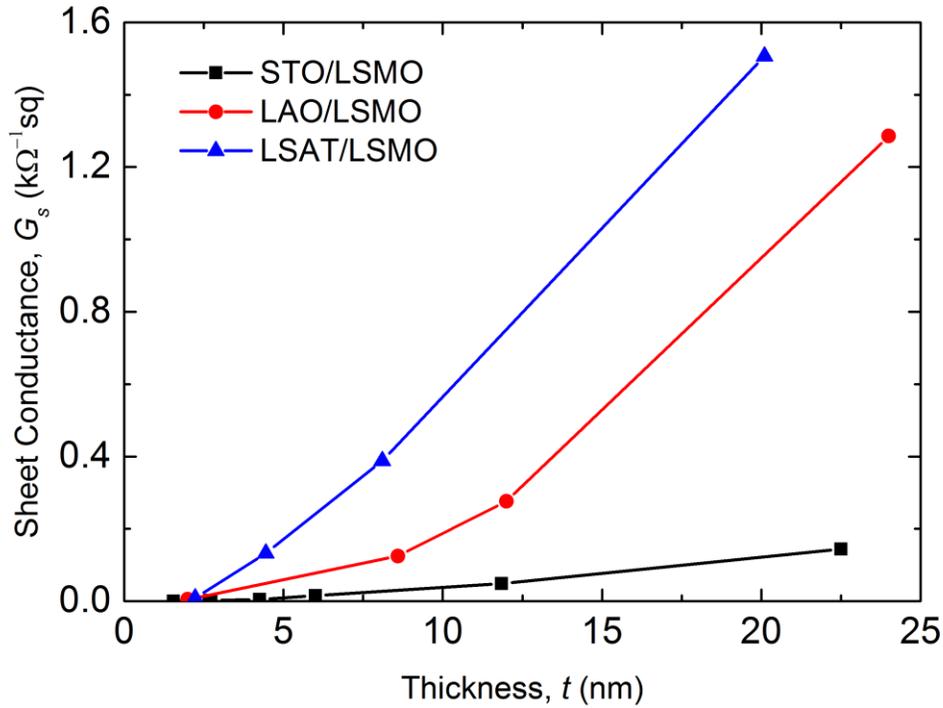

Fig 5. Plot to estimate the electrical dead-layer thicknesses of LSMO on different substrates.

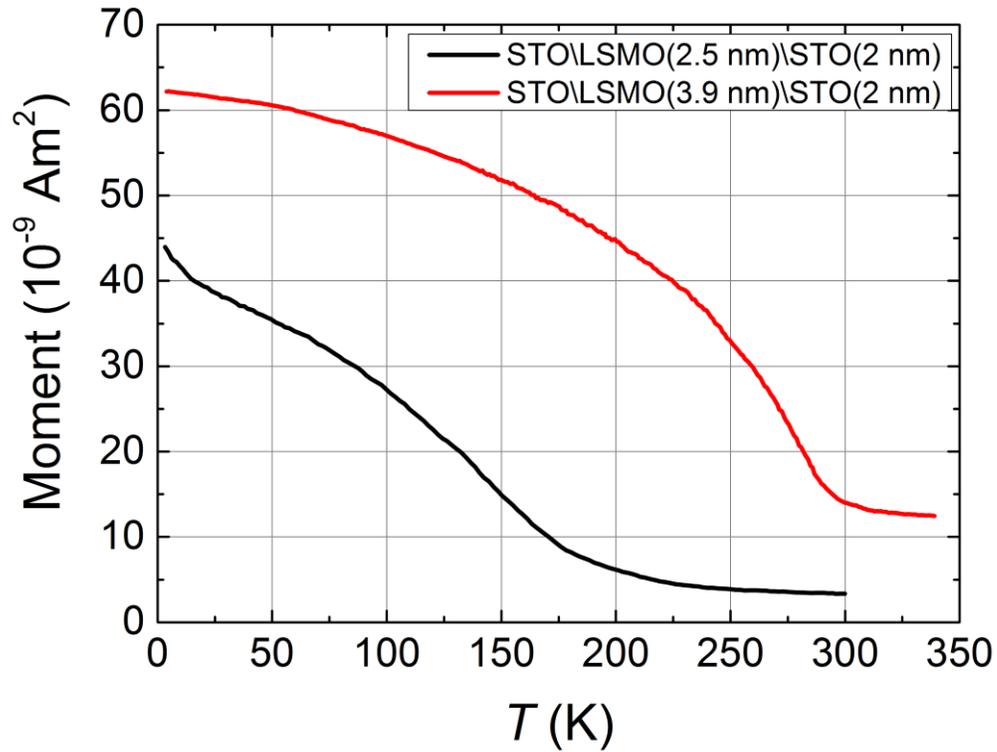

Fig 6. Thermal scans in 100 mT of two thin films of LSMO on STO, covered by STO cap layers.